\documentclass[a4paper,twocolumn,english,prl, twocolumn, superscriptaddress]{revtex4}
\pdfoutput=1
\usepackage[latin9]{inputenc}
\usepackage{float}
\usepackage{graphicx}

\makeatletter

\usepackage{babel}
\makeatother

\begin{document}

\title{Self-induced oscillations in an optomechanical system}

\author{Max Ludwig{*}}

\address{Arnold-Sommerfeld Center for Theoretical Physics}

\address{Center for NanoScience and Department of Physics, Ludwig-Maximilians
Universit\"at M\"unchen, Munich, Germany}

\author{Clemens Neuenhahn{*}}

\address{Arnold-Sommerfeld Center for Theoretical Physics}

\address{Center for NanoScience and Department of Physics, Ludwig-Maximilians
Universit\"at M\"unchen, Munich, Germany}

\author{Constanze Metzger}

\address{Boston University, Electrical and Computer Engineering, Boston, MA
02215, USA}

\address{Center for NanoScience and Department of Physics, Ludwig-Maximilians
Universit\"at M\"unchen, Munich, Germany}

\author{Alexander Ortlieb}

\address{Center for NanoScience and Department of Physics, Ludwig-Maximilians
Universit\"at M\"unchen, Munich, Germany}

\author{Ivan Favero}

\address{Laboratoire Matériaux et Phénomènes Quantiques, Université Paris
Diderot et CNRS, UMR 7162, Bâtiment Condorcet, 75205 Paris Cedex 13,
France}

\address{Center for NanoScience and Department of Physics, Ludwig-Maximilians
Universit\"at M\"unchen, Munich, Germany}

\author{Khaled Karrai}

\address{Center for NanoScience and Department of Physics, Ludwig-Maximilians
Universit\"at M\"unchen, Munich, Germany}

\author{Florian Marquardt}

\address{Arnold-Sommerfeld Center for Theoretical Physics}

\address{Center for NanoScience and Department of Physics, Ludwig-Maximilians
Universit\"at M\"unchen, Munich, Germany}

\begin{abstract}
We have explored the nonlinear dynamics of an optomechanical system
consisting of an illuminated Fabry-Perot cavity, one of whose end-mirrors
is attached to a vibrating cantilever. Such a system can experience
negative light-induced damping and enter a regime of self-induced
oscillations. We present a systematic experimental and theoretical
study of the ensuing attractor diagram describing the nonlinear dynamics,
in an experimental setup where the oscillation amplitude becomes large,
and the mirror motion is influenced by several optical modes. A theory
has been developed that yields detailed quantitative agreement with
experimental results. This includes the observation of a regime where
two mechanical modes of the cantilever are excited simultaneously.
\end{abstract}
\maketitle
Micro- and nanomechanical systems have become a focus of both theoretical
and experimental research \citep{2005_07_SchwabRoukes_NEMS}, with
the goals ranging from ultrasensitive measurements to fundamental
tests of quantum mechanics. One particularly promising branch of this
field deals with optomechanical systems, where the interaction of
light (stored inside an optical cavity) with macroscopic mechanical
degrees of freedom (such as the coordinate of a cantilever) is exploited.
This can give rise to a variety of effects, including a modification
of the mechanical spring constant \citep{1967_BraginskyManukin_PonderomotiveEffectsEMRadiation,2003_08_Vogel_PhotothermalForceOnCantilever,2004_12_HoehbergerKarrai_CoolingMicroleverNature,2006_07_Arcizet_CoolingMirror,2006_12_NergisMavalvala_LIGO},
bistability \citep{1983_10_DorselWalther_BistabilityMirror,1985_11_Meystre_RadiationPressureDrivenInterferometers},
optomechanical cooling, and parametric instability. A recent series
of experiments has demonstrated impressive progress with respect to
cooling \citep{2004_12_ConstanzeKhaled_WithNote,2006_11_Bouwmeester_FeedbackCooling,2006_07_Arcizet_CoolingMirror,2006_05_AspelmeyerZeilinger_SelfCoolingMirror,2006_11_Kippenberg_RadPressureCooling,2006_12_NergisMavalvala_LIGO,2007_Favero_OpticalCoolingMicromirror,2007_07_Harris_MembraneInTheMiddle},
which may ultimately lead to the quantum ground state \citep{2007_01_Marquardt_CantileverCooling,2007_02_WilsonRae_Cooling}
of mechanical motion in such devices. On the other hand, the opposite
regime is of equal interest, where the mechanical $Q$ factor is enhanced
since the mechanical degree of freedom extracts energy provided by
the optical radiation. In that regime, a parametric instability arises
which drives the system into a state of self-sustained oscillations
\citep{1967_BraginskyManukin_PonderomotiveEffectsEMRadiation,1970_Braginsky_OpticalCoolingExperiment,2001_07_Braginsky_ParametricInstabilityFPCavity,2002_KimLee_SelfoscillationsAFM,2004_KarraiConstanze_IEEE,2005_06_Vahala_SelfOscillationsCavity,2005_07_VahalaTheoryPRL,2005_02_MarquardtHarrisGirvin_Cavity,2006_08_Corbitt_InstabilityLIGO,2007_CarmonVahala_ModeSpectroscopy}.
Moreover, it is now known that the physics of that regime also applies
to other systems as diverse as an LC circuit driven by a radio-frequency
source \citep{2007_Wineland_RFcircuitCooling} or a current-biased
superconducting single-electron transistor coupled to a nanobeam \citep{2006_08_Schwab_CPB_Molasses,2007_Rodrigues_InstabilitySSET}.
Although the basic instability has been observed by now in a number
of experiments \citep{2002_KimLee_SelfoscillationsAFM,2004_KarraiConstanze_IEEE,2005_06_Vahala_SelfOscillationsCavity,2005_07_VahalaTheoryPRL,2006_08_Corbitt_InstabilityLIGO,2007_CarmonVahala_ModeSpectroscopy},
it was recently realized theoretically \citep{2005_02_MarquardtHarrisGirvin_Cavity}
that the nonlinear dynamics of this system can become highly nontrivial,
leading to an intricate attractor diagram. Here we report on an experiment
that traces this diagram, and combine it with a detailed theoretical
analysis and systematic quantitative comparison of theory and experiment.
In addition, this comparison has enabled us to observe an unexpected
feature, the simultaneous excitation of several mechanical modes of
the cantilever, leading to coupled nonlinear dynamics. %
\begin{figure}
\includegraphics[bb=0bp 0bp 237bp 178bp,clip,width=1\columnwidth]{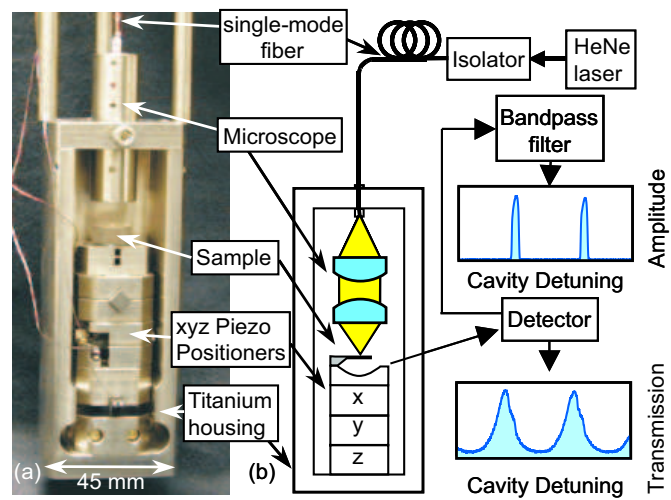}

\caption{\label{fig:setup}(Color online) The experimental setup. The light
inside the optical cavity is focused onto a cantilever, where it exerts
a force. Both the average transmitted light and its sidebands at the
cantilever frequency are recorded ({}``Transmission'' and {}``Amplitude'',
respectively).}

\end{figure}

\emph{Experimental setup}. - To study these questions, we employ the
optomechanical setup displayed in Fig.~\ref{fig:setup}. The light
of a $633\,{\rm nm}$ monomode HeNe-laser is coupled into a single
mode fiber and passes through a Faraday isolator (35 dB suppression).
The fiber end inside a vacuum chamber (at $5\cdot10^{-6}\,{\rm mbar}$)
was polished and coated with a reflecting gold layer of $30\,{\rm nm}$
(yielding a theoretical reflectivity of $70\%$) to form the first
cavity mirror. The sample is a gold-coated AFM cantilever acting as
a micromirror, with length 223 $\mu$m, thickness 470 nm, width 22
$\mu$m, spring constant $K=0.01\,{\rm N/m}$, and a gold layer of
$42\,{\rm nm}$ evaporated on one side only. The cantilever's fundamental
mechanical mode has a frequency of $\omega_{1}=2\pi\cdot8.7\,{\rm kHz}$
and a damping rate of $\Gamma_{1}=30\,{\rm Hz}$. A simulation of
the silicon-gold bilayer system gave a reflectivity of 91$\%$ for
a wavelength of 633 nm. The divergent beam coming out of the fiber
is sent through a microscope setup consisting of two identical lenses,
yielding a Gaussian focus on the sample with a $1/e^{2}$-diameter
of $6\,\mu{\rm m}$. The cantilever has been mounted on an $xyz$
piezo stepper positioner block %
\footnote{xyz positioner from attocube systems AG (Munich)%
}, such that it can be placed at the microscope's focal point, which
was chosen near the end of the cantilever, at about 3/4 of its length.
The finesse of the cavity defined by the sample and the fiber end
was found to be $F\approx4.5$. The transmitted intensity is measured
with a Si photodiode behind the cantilever, while sweeping the cantilever
position through the optical resonance. 

\emph{Theoretical model.} - The dynamics of the cantilever is described
by the equation of motion of a damped oscillator, driven by light-induced
forces: \begin{equation}
\ddot{x}=-\omega_{1}^{2}(x-x_{0})-\Gamma_{1}\dot{x}+(F^{{\rm rad}}+F^{{\rm bol}})/m_{1}\label{eq:eo}\end{equation}
Here $x(t)$ is the cantilever deflection observed at the laser spot.
For now, we focus on the motion of the first mechanical mode (with
effective mass $m_{1}$, frequency $\omega_{1}$, and damping rate
$\Gamma_{1}$), and we will return to the influence of higher-order
modes below. The cavity is assumed to be in resonance with the laser
when $x=0$, while the mechanical equilibrium position is given by
$x_{0}$ (sometimes referred to as the 'detuning'). In the experiment,
it is controlled by the piezo positioning system. The radiation pressure
force $F^{{\rm rad}}$ is proportional to the power $I$ circulating
inside the cavity, $F^{{\rm rad}}/m_{1}=\mathcal{P}\cdot I$. The
bolometric force $F^{{\rm bol}}$ arises due to light being absorbed,
thus heating the cantilever that then deforms as a bimorph. It is
enhanced by a factor $\Lambda$ over $F^{{\rm rad}}$ and is retarded
due to the finite time of thermal conductance $\tau$:\begin{equation}
F^{{\rm bol}}/m_{1}=\Lambda\mathcal{P}\int_{-\infty}^{t}\frac{dt'}{\tau}e^{-(t-t')/\tau}I(t')\equiv\Lambda\mathcal{P}\theta(t).\end{equation}
Here $\theta$ is proportional to the change in temperature brought
about by absorption of light.

We restrict the following discussion to the case of an optical ring
down time that is much smaller than the period of cantilever oscillations
(like in the present setup). Thus, the light intensity reacts instantaneously
to the cantilever motion, $I(t)=I[x(t)]$. Due to the small finesse,
we have to employ the Airy function dependence describing a series
of overlapping Fabry-Perot resonances:\begin{equation}
\frac{I[x(t)]}{I_{{\rm max}}}=\frac{1}{1+(2F/\pi)^{2}\sin(\frac{2\pi}{\lambda}x(t))^{2}}.\label{eq:intensity}\end{equation}
Here $I_{{\rm max}}$ is the peak circulating power, which is proportional
to the input power. In contrast to the theory in \citep{2005_02_MarquardtHarrisGirvin_Cavity},
where a large optical ringdown time led to intensity oscillations,
here the nonlinear dynamics is induced by the time retardation of
the bolometric force.

\emph{Self-induced oscillations}. - Time-retarded forces induce an
effective optomechanical damping rate whose sign changes when passing
through the resonance \citep{1967_BraginskyManukin_PonderomotiveEffectsEMRadiation,1970_Braginsky_OpticalCoolingExperiment,2004_12_HoehbergerKarrai_CoolingMicroleverNature,2005_02_MarquardtHarrisGirvin_Cavity,2007_01_Marquardt_CantileverCooling,2007_02_WilsonRae_Cooling}.
When the full damping rate becomes negative, the system undergoes
a Hopf bifurcation and the cantilever settles into stable oscillations
\citep{2005_02_MarquardtHarrisGirvin_Cavity} which (for the parameters
of interest here) are sinusoidal to a very good approximation: $x(t)=\bar{x}+A\cos(\omega_{1}t)$.
The nonlinear dynamics can then be characterized by solving for the
amplitude $A$ and offset $\bar{x}$. From these, it will be possible
to obtain the experimentally observed evolution of the light intensity
$I(t)$. In steady state, the average force and power input must be
zero, i.e. $\left\langle \ddot{x}\right\rangle =0$ and $\left\langle \ddot{x}\dot{x}\right\rangle =0$,
where $\left\langle \ldots\right\rangle $ denotes the time-average.
Inserting Eq.~(\ref{eq:eo}), we obtain the power balance equation:
\begin{equation}
\mathcal{P}\left\langle \dot{x}(I(t)+\Lambda\theta(t))\right\rangle =\Gamma_{1}\left\langle \dot{x}^{2}\right\rangle \,.\label{eq:powero}\end{equation}
The radiation pressure does not contribute, $\left\langle \dot{x}I\right\rangle =0$,
since the intensity follows the motion instantaneously (thus $\dot{x}I$
is an antisymmetric function of time). The expression $\left\langle \dot{x}\theta\right\rangle $
may be simplified by introducing the first harmonic of the light intensity,
$\tilde{I}_{1}=\frac{1}{T}\int_{0}^{T}dt\,\cos(\omega_{1}t)I(t)$
(with the period of mechanical motion $T=2\pi/\omega_{1}$):\begin{equation}
\left\langle \dot{x}\theta\right\rangle =-A\omega_{1}\cdot\tilde{I}_{1}\frac{\omega_{1}\tau}{(\omega_{1}\tau)^{2}+1}\,.\end{equation}
 This yields the first of the equations needed to find $\left(\bar{x},A\right)$:
\begin{equation}
\frac{\Gamma_{1}}{\mathcal{P}}=\frac{-\Lambda}{A\omega_{1}\pi}\frac{\omega_{1}\tau}{(\omega_{1}\tau)^{2}+1}\int_{0}^{2\pi}d\varphi\, I[\bar{x}+A\cos\varphi]\cos\varphi\label{eq:powert}\end{equation}
 Moreover, the force balance condition\begin{equation}
\omega_{1}^{2}(\bar{x}-x_{0})=\mathcal{P}\cdot\left(\left\langle I\right\rangle +\Lambda\left\langle \theta\right\rangle \right)\label{eq:forceo}\end{equation}
 enables us to obtain $\bar{x}=\bar{x}(x_{0},A)$: \begin{equation}
\bar{x}-x_{0}=\frac{(1+\Lambda)\mathcal{P}}{2\pi\omega_{1}^{2}}\int_{0}^{2\pi}d\varphi\, I[\bar{x}+A\cos\varphi].\label{eq:forcet}\end{equation}
As a special case, for $A=0$, this contains the physics of static
optomechanical bistability \citep{1983_10_DorselWalther_BistabilityMirror,1985_11_Meystre_RadiationPressureDrivenInterferometers}.
\begin{figure}[H]
\includegraphics[bb=0bp 0bp 400bp 641bp,clip,width=1\columnwidth]{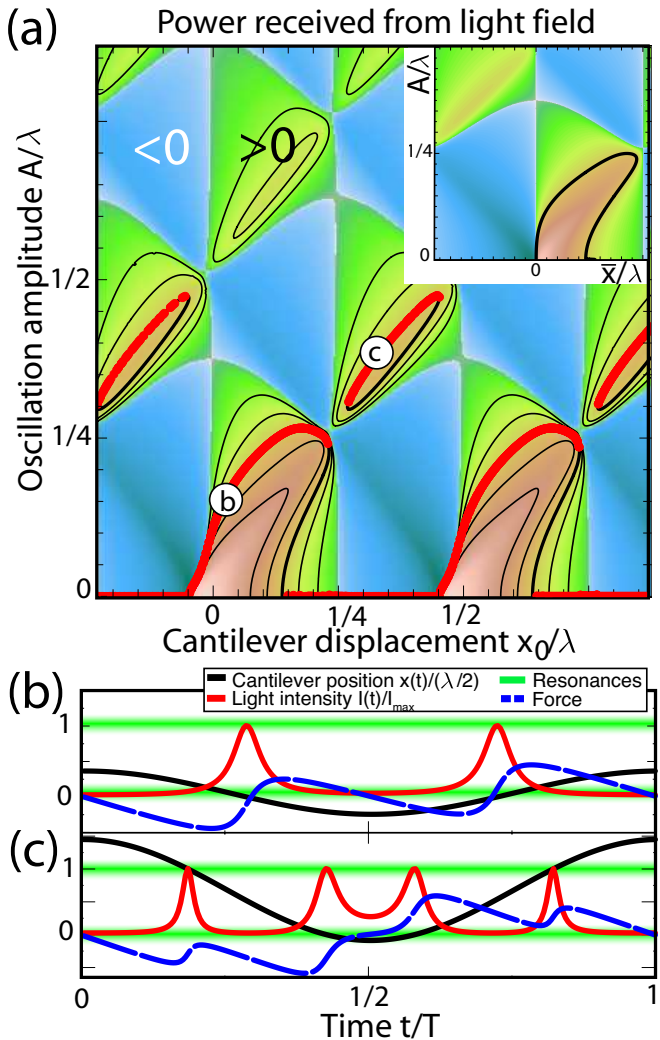}

\caption{\label{fig:atdiagram}(Color online). Theoretically predicted attractor
diagram for the nonlinear cantilever dynamics. (a) Amplitude $A$
of optomechanically induced oscillations vs. equilibrium position
$x_{0}$ of the cantilever. The colorscale plot displays the net power
fed into the cantilever from the radiation field, i.e. the r.h.s.
of Eq.~\ref{eq:powert}. Contour lines indicate possible oscillation
amplitudes (dynamical attractors) for different values of the damping
constant $\Gamma_{1}$. Red dots are the results of direct numerical
simulations, for the value of $\Gamma_{1}$ realized in the present
experimental setup (and a laser input power level of $I=0.3I_{0}$).
The inset displays the same plot, but as a function of $(A,\bar{x})$
instead of $(A,x_{0})$, i.e. without the distortion due to the time-averaged
radiation force that leads to $\bar{x}\neq x_{0}$. (b,c) Whenever
the cantilever motion (thick black line) passes through the optical
resonances (green bars), the light intensity (red line) displays spikes,
which lead to delayed increases in the radiation force (dashed blue
line), which was plotted as $(\theta-\bar{\theta})\times200$. Plots
(b) and (c) correspond to the respective positions indicated in the
attractor diagram (a).}

\end{figure}

\emph{Attractor diagram}. - Using Eqs.~(\ref{eq:powert}) and (\ref{eq:forcet}),
one obtains solutions $(\bar{x},A)$ which can be visualized in attractor
diagrams, like the one shown in Fig.~\ref{fig:atdiagram}. The color
scale encodes the power input due to the light-induced forces (r.h.s.
of Eq.~\ref{eq:powert}), as a function of $x_{0}$ and $A$. The
solution of Eq.~\ref{eq:powert} for various values of $\Gamma_{1}/\mathcal{P}$
then corresponds to contour lines of this function. Apart from the
expected $\lambda/2$-periodicity in the detuning $x_{0}$, the main
feature is the appearance of multiple solutions for $A$ at a given
$x_{0}$ ({}``dynamical multistability''). Self-induced oscillations
appear when $x(t)$ briefly dips into the resonance near its turning
point, thereby gaining energy. Thus, to a first approximation, attractors
appear along the diagonals $A\approx\bar{x}+n\lambda/2$. Near the
special points $\bar{x}=\lambda/4+n\lambda/2$, the total power input
vanishes due to symmetry: The cantilever extracts energy from the
light field at one turning point but loses an equal amount of energy
at the other turning point.

The deviation between $\bar{x}$ and $x_{0}$, obtained from Eq. (\ref{eq:forcet}),
leads to a distortion of the diagram (compare inset of Fig.~\ref{fig:atdiagram}).
This effect grows with increasing input power, finally leading to
multiple solutions for $\bar{x}(x_{0},A)$. 

\emph{Comparison of theory and experiment}. - In the experiment, the
detuning $x_{0}$ and the input power $I_{{\rm in}}$ are varied,
while the transmitted light intensity is measured. This can be compared
to the time-averaged circulating power $\left\langle I(t)\right\rangle $
obtained from the theory. Moreover, as soon as the self-induced oscillations
set in, $I(t)$ is modulated at the cantilever frequency. A very helpful
nontrivial feature of this system is the existence of a relation between
the motional amplitude $A$ and the first harmonic of the light intensity,
$\tilde{I}_{1}$ (defined above). From Eq.~(\ref{eq:powert}), we
see that they are directly proportional, with the proportionality
factor containing only known, fixed parameters: $A=-2(\Lambda\mathcal{P}/\omega_{1}\Gamma)\omega_{1}\tau(1+(\omega_{1}\tau)^{2})^{-1}\tilde{I}_{1}$.
This relation holds true only in steady state (on the attractors),
but then it is valid even when the motion sweeps across several optical
resonances. Experimentally, the first harmonic $\tilde{I}_{1}$ is
obtained by sending the photodetector signal through a narrow bandpass
filter ($100\,{\rm Hz}$) centered at the eigenfrequency $\omega_{1}$
of the first mechanical mode. 

\begin{figure}
\begin{centering}
\includegraphics[bb=75bp 0bp 386bp 433bp,clip,width=1\columnwidth]{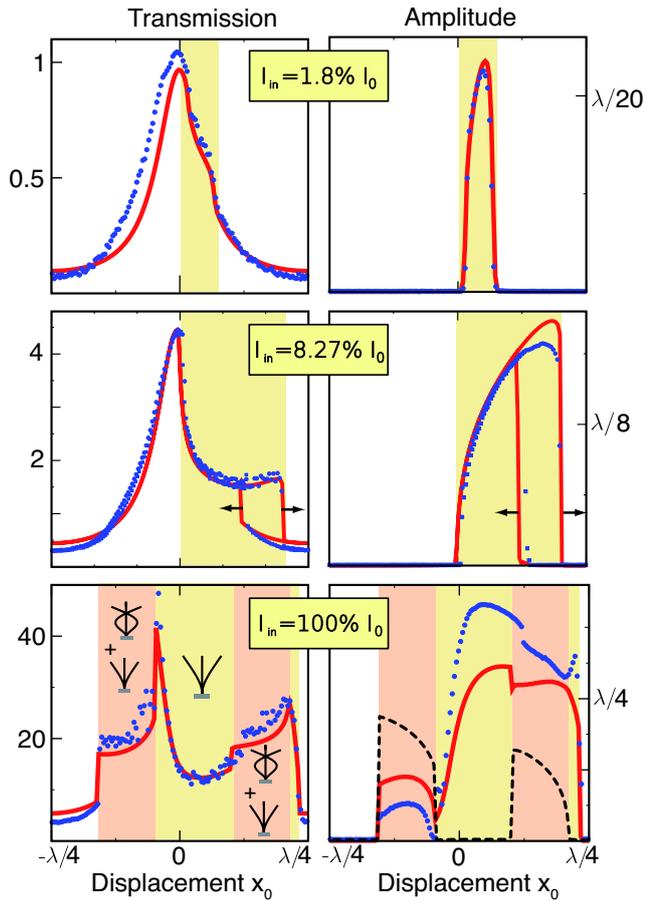}
\par\end{centering}

\caption{(Color online) Experiment vs. theory. The transmitted light intensity
(left) and the amplitude of self-induced cantilever oscillations (right),
from a simulation of the theoretical model (red full curves) and from
the experiment (blue data points), at increasing input power levels
(top to bottom). Theoretical {}``transmission'' curves display the
(rescaled) time-averaged circulating light intensity from the simulation.
{}``Amplitude'' curves are obtained from the power in the intensity
sidebands at the fundamental mechanical mode of the cantilever (see
main text). For clarity, the hysteresis observed upon sweeping $x_{0}$
up or down has been shown only in the middle panel. The region of
instability (shaded interval) grows with increasing input power. Simultaneous
self-induced oscillations of the first two mechanical modes set in
at the highest power displayed (in the two intervals indicated in
the plot). The calculated amplitude of the second mode is shown as
a dashed line. \label{fig:gesamt}}

\end{figure}

Theoretical and experimental curves for both the average intensity
({}``transmission'') and the amplitude are shown in Fig.~\ref{fig:gesamt},
for different input powers. For this comparison, the following parameters
have been used: $F\approx4.5$ (from a fit at low input power), $\Lambda=3950$
and $\omega_{1}\tau=39$ (obtained independently). The overall conversion
factor between experimentally measured input power and the force on
the cantilever was found to be $\mathcal{P}I_{{\rm 0}}=0.0775{\rm m/s^{2}}$,
by adjusting for a good fit to the data at intermediate power (the
same was done for the rescaling of theoretical and experimental transmission
intensity). Here the maximum laser power $I_{0}=1.3\,{\rm mW}$ is
estimated to yield $500\,\mu{\rm W}$ circulating in the cavity on
resonance. 

At the lowest power displayed in Fig.~\ref{fig:gesamt}, self-oscillations
have just set in, and the transmission curve shows a striking asymmetry.
At higher input powers, the multistability predicted by the attractor
diagram, Fig.~\ref{fig:atdiagram}, leads to hysteresis effects that
appear upon sweeping $x_{0}$ up and down.

Beginning at $I_{{\rm in}}=0.57\, I_{0}$, a second interval of self-oscillatory
behaviour appears to the \emph{left} of the resonance, growing stronger
and wider with increasing laser power. This initially completely unexpected
result may be explained by invoking the influence of higher mechanical
modes. These may be excited by the radiation as well, leading to coupled
(multimode) nonlinear dynamics with richer features than discussed
up to now. %

In order to describe the behaviour in that case, we now take into
account the second mode as well. The total displacement is $x(t)=x_{0}+x_{1}(t)+x_{2}(t)$,
and we have to employ a set of equations: \begin{equation}
\ddot{x}_{i}=-\omega_{i}^{2}x_{i}-\Gamma_{i}\dot{x_{i}}+F_{i}^{{\rm bol}}[x(t)]/m_{i}\,,\end{equation}
where $x_{i}$ denotes the coordinate of the $i$-th mechanical mode
with frequency $\omega_{i}$, mechanical damping rate $\Gamma_{i}$,
and effective mass $m_{i}$ (where $\omega_{1}/2\pi=8.7\,{\rm kHz}$,
$\omega_{2}/2\pi=60\,{\rm kHz}$, $\Gamma_{1}=30.0\,{\rm Hz}$, $\Gamma_{2}=150\,{\rm Hz}$).
We neglected the radiation pressure force, as this is much smaller
anyway for the parameters of this setup. The mechanical modes are
now coupled indirectly by the bolometric force. For the present setup,
this force changes sign when going to the second mode. Choosing $F_{2}^{{\rm bol}}m_{1}/F_{1}^{{\rm bol}}m_{2}=-28.8$
as an adjustable parameter, we have found the numerical simulation
of these coupled nonlinear equations for the first two modes to be
in surprisingly good agreement with the experiment (Fig.~\ref{fig:gesamt}).
We have to note that the relation between the measured {}``amplitude'',
i.e. first harmonic of $I(t)$ at frequency $\omega_{1}$, and the
actual amplitude $A_{1}$ does not hold exactly if both modes are
excited simultaneously. 

At maximum laser power, there are actually two intervals with simultaneous
excitation of both modes (indicated in Fig.~\ref{fig:gesamt}). Specifically,
the onset of such a regime at $x_{0}\approx\lambda/8$ can be interpreted
as follows: Taking into account $F_{2}^{{\rm bol}}/F_{1}^{{\rm bol}}<0$,
we see that the second mode gains its energy from dipping into the
resonance at $x=\lambda/2$, while the first is still provided with
energy due to the resonance at $x=0$.

Numerical evidence shows that the steady-state motion consists of
sinusoidal oscillations in $x_{1,2}$ at the respective eigenfrequencies,
of nearly constant amplitudes and without phase locking (for the parameters
explored here). Thus $x(t)\approx x_{0}+\sum_{i=1}^{2}[A_{i}+\delta A_{i}(t)]\cos(\omega_{i}t+\phi_{i})$,
where $\delta A_{i}(t)/A_{i}\ll1$, and the $\phi_{i}$ are arbitrary
phases. Higher input powers will lead to excitations of additional
modes, and the system might go into a chaotic regime of motion.

\emph{Conclusions}. - We have analyzed the nonlinear dynamics of an
optomechanical system, by measuring and explaining its attractor diagram.
The comparison of data and theoretical predictions have revealed the
onset of multi-mode dynamics at large optical power, with two mechanical
modes of the cantilever participating in the radiation-driven self-sustained
oscillations. These effects could find applications in highly sensitive
force or displacement detection \citep{2005_02_MarquardtHarrisGirvin_Cavity}.
In the future, it would be interesting to observe the attractor diagram
in systems of a high optical finesse \citep{2005_06_Vahala_SelfOscillationsCavity,2005_07_VahalaTheoryPRL}
(with delayed radiation dynamics), or the self-excitation of multiple
mechanical modes of sub-wavelength mechanical resonators interacting
with the radiation field inside a cavity \citep{2007_07_FaveroKarrai_ScatteringCooling,2007_07_Harris_MembraneInTheMiddle}.
The whole field of quantum nonlinear dynamics in systems of this kind
also remains to be explored.

We acknowledge support by the Nanosystems Initiative Munich (NIM).
F. M. acknowledges support by an Emmy-Noether grant of the DFG, and
I.~F. acknowledges the A.~v.~Humboldt Foundation. 

{*} Note: The first two authors (M.~L. and C.~N.) contributed equally
to this work.

\bibliographystyle{apsrev}
\bibliography{/Users/florian/pre/bib/BibFM}

\begin{thebibliography}{29}
\expandafter\ifx\csname natexlab\endcsname\relax\def\natexlab#1{#1}\fi
\expandafter\ifx\csname bibnamefont\endcsname\relax
  \def\bibnamefont#1{#1}\fi
\expandafter\ifx\csname bibfnamefont\endcsname\relax
  \def\bibfnamefont#1{#1}\fi
\expandafter\ifx\csname citenamefont\endcsname\relax
  \def\citenamefont#1{#1}\fi
\expandafter\ifx\csname url\endcsname\relax
  \def\url#1{\texttt{#1}}\fi
\expandafter\ifx\csname urlprefix\endcsname\relax\def\urlprefix{URL }\fi
\providecommand{\bibinfo}[2]{#2}
\providecommand{\eprint}[2][]{\url{#2}}

\bibitem[{\citenamefont{Schwab and Roukes}(2005)}]{2005_07_SchwabRoukes_NEMS}
\bibinfo{author}{\bibfnamefont{K.~C.} \bibnamefont{Schwab}} \bibnamefont{and}
  \bibinfo{author}{\bibfnamefont{M.~L.} \bibnamefont{Roukes}},
  \bibinfo{journal}{Physics Today} \textbf{\bibinfo{volume}{July Issue}},
  \bibinfo{pages}{36} (\bibinfo{year}{2005}).

\bibitem[{\citenamefont{Braginsky and
  Manukin}(1967)}]{1967_BraginskyManukin_PonderomotiveEffectsEMRadiation}
\bibinfo{author}{\bibfnamefont{V.}~\bibnamefont{Braginsky}} \bibnamefont{and}
  \bibinfo{author}{\bibfnamefont{A.}~\bibnamefont{Manukin}},
  \bibinfo{journal}{Soviet Physics JETP} \textbf{\bibinfo{volume}{25}},
  \bibinfo{pages}{653} (\bibinfo{year}{1967}).

\bibitem[{\citenamefont{Vogel et~al.}(2003)\citenamefont{Vogel, Mooser, Karrai,
  and Warburton}}]{2003_08_Vogel_PhotothermalForceOnCantilever}
\bibinfo{author}{\bibfnamefont{M.}~\bibnamefont{Vogel}},
  \bibinfo{author}{\bibfnamefont{C.}~\bibnamefont{Mooser}},
  \bibinfo{author}{\bibfnamefont{K.}~\bibnamefont{Karrai}}, \bibnamefont{and}
  \bibinfo{author}{\bibfnamefont{R.~J.} \bibnamefont{Warburton}},
  \bibinfo{journal}{Applied Physics Letters} \textbf{\bibinfo{volume}{83}},
  \bibinfo{pages}{1337} (\bibinfo{year}{2003}).

\bibitem[{\citenamefont{{H\"ohberger}-Metzger and
  Karrai}(2004{\natexlab{a}})}]{2004_12_HoehbergerKarrai_CoolingMicroleverNatu%
re}
\bibinfo{author}{\bibfnamefont{C.}~\bibnamefont{{H\"ohberger}-Metzger}}
  \bibnamefont{and} \bibinfo{author}{\bibfnamefont{K.}~\bibnamefont{Karrai}},
  \bibinfo{journal}{Nature} \textbf{\bibinfo{volume}{432}},
  \bibinfo{pages}{1002} (\bibinfo{year}{2004}{\natexlab{a}}).

\bibitem[{\citenamefont{Arcizet et~al.}(2006)\citenamefont{Arcizet, Cohadon,
  Briant, Pinard, and Heidmann}}]{2006_07_Arcizet_CoolingMirror}
\bibinfo{author}{\bibfnamefont{O.}~\bibnamefont{Arcizet}},
  \bibinfo{author}{\bibfnamefont{P.~F.} \bibnamefont{Cohadon}},
  \bibinfo{author}{\bibfnamefont{T.}~\bibnamefont{Briant}},
  \bibinfo{author}{\bibfnamefont{M.}~\bibnamefont{Pinard}}, \bibnamefont{and}
  \bibinfo{author}{\bibfnamefont{A.}~\bibnamefont{Heidmann}},
  \bibinfo{journal}{Nature} \textbf{\bibinfo{volume}{444}}, \bibinfo{pages}{71}
  (\bibinfo{year}{2006}).

\bibitem[{\citenamefont{Corbitt et~al.}(2007)}]{2006_12_NergisMavalvala_LIGO}
\bibinfo{author}{\bibfnamefont{T.}~\bibnamefont{Corbitt}} \bibnamefont{et~al.},
  \bibinfo{journal}{\prl} \textbf{\bibinfo{volume}{98}},
  \bibinfo{pages}{150802} (\bibinfo{year}{2007}).

\bibitem[{\citenamefont{Dorsel et~al.}(1983)\citenamefont{Dorsel, McCullen,
  Meystre, Vignes, and Walther}}]{1983_10_DorselWalther_BistabilityMirror}
\bibinfo{author}{\bibfnamefont{A.}~\bibnamefont{Dorsel}},
  \bibinfo{author}{\bibfnamefont{J.~D.} \bibnamefont{McCullen}},
  \bibinfo{author}{\bibfnamefont{P.}~\bibnamefont{Meystre}},
  \bibinfo{author}{\bibfnamefont{E.}~\bibnamefont{Vignes}}, \bibnamefont{and}
  \bibinfo{author}{\bibfnamefont{H.}~\bibnamefont{Walther}},
  \bibinfo{journal}{\prl} \textbf{\bibinfo{volume}{51}}, \bibinfo{pages}{1550}
  (\bibinfo{year}{1983}).

\bibitem[{\citenamefont{Meystre et~al.}(1985)\citenamefont{Meystre, Wright,
  McCullen, and
  Vignes}}]{1985_11_Meystre_RadiationPressureDrivenInterferometers}
\bibinfo{author}{\bibfnamefont{P.}~\bibnamefont{Meystre}},
  \bibinfo{author}{\bibfnamefont{E.~M.} \bibnamefont{Wright}},
  \bibinfo{author}{\bibfnamefont{J.~D.} \bibnamefont{McCullen}},
  \bibnamefont{and} \bibinfo{author}{\bibfnamefont{E.}~\bibnamefont{Vignes}},
  \bibinfo{journal}{J. Opt. Soc. Am. B} \textbf{\bibinfo{volume}{2}},
  \bibinfo{pages}{1830} (\bibinfo{year}{1985}).

\bibitem[{\citenamefont{{H\"ohberger}-Metzger and
  Karrai}(2004{\natexlab{b}})}]{2004_12_ConstanzeKhaled_WithNote}
\bibinfo{author}{\bibfnamefont{C.}~\bibnamefont{{H\"ohberger}-Metzger}}
  \bibnamefont{and} \bibinfo{author}{\bibfnamefont{K.}~\bibnamefont{Karrai}},
  \bibinfo{journal}{Nature} \textbf{\bibinfo{volume}{432}},
  \bibinfo{pages}{1002} (\bibinfo{year}{2004}{\natexlab{b}}).

\bibitem[{\citenamefont{Kleckner and
  Bouwmeester}(2006)}]{2006_11_Bouwmeester_FeedbackCooling}
\bibinfo{author}{\bibfnamefont{D.}~\bibnamefont{Kleckner}} \bibnamefont{and}
  \bibinfo{author}{\bibfnamefont{D.}~\bibnamefont{Bouwmeester}},
  \bibinfo{journal}{Nature} \textbf{\bibinfo{volume}{444}}, \bibinfo{pages}{75}
  (\bibinfo{year}{2006}).

\bibitem[{\citenamefont{Gigan
  et~al.}(2006)}]{2006_05_AspelmeyerZeilinger_SelfCoolingMirror}
\bibinfo{author}{\bibfnamefont{S.}~\bibnamefont{Gigan}} \bibnamefont{et~al.},
  \bibinfo{journal}{Nature} \textbf{\bibinfo{volume}{444}}, \bibinfo{pages}{67}
  (\bibinfo{year}{2006}).

\bibitem[{\citenamefont{Schliesser et~al.}(2006)\citenamefont{Schliesser,
  Del'Haye, Nooshi, Vahala, and
  Kippenberg}}]{2006_11_Kippenberg_RadPressureCooling}
\bibinfo{author}{\bibfnamefont{A.}~\bibnamefont{Schliesser}},
  \bibinfo{author}{\bibfnamefont{P.}~\bibnamefont{Del'Haye}},
  \bibinfo{author}{\bibfnamefont{N.}~\bibnamefont{Nooshi}},
  \bibinfo{author}{\bibfnamefont{K.~J.} \bibnamefont{Vahala}},
  \bibnamefont{and} \bibinfo{author}{\bibfnamefont{T.~J.}
  \bibnamefont{Kippenberg}}, \bibinfo{journal}{\prl}
  \textbf{\bibinfo{volume}{97}}, \bibinfo{pages}{243905}
  (\bibinfo{year}{2006}).

\bibitem[{\citenamefont{Favero
  et~al.}(2007)}]{2007_Favero_OpticalCoolingMicromirror}
\bibinfo{author}{\bibfnamefont{I.}~\bibnamefont{Favero}} \bibnamefont{et~al.},
  \bibinfo{journal}{Applied Physics Letters} \textbf{\bibinfo{volume}{90}},
  \bibinfo{pages}{104101} (\bibinfo{year}{2007}).

\bibitem[{\citenamefont{Thompson et~al.}(2007)\citenamefont{Thompson, Zwickl,
  Jayich, Marquardt, Girvin, and Harris}}]{2007_07_Harris_MembraneInTheMiddle}
\bibinfo{author}{\bibfnamefont{J.~D.} \bibnamefont{Thompson}},
  \bibinfo{author}{\bibfnamefont{B.~M.} \bibnamefont{Zwickl}},
  \bibinfo{author}{\bibfnamefont{A.~M.} \bibnamefont{Jayich}},
  \bibinfo{author}{\bibfnamefont{F.}~\bibnamefont{Marquardt}},
  \bibinfo{author}{\bibfnamefont{S.~M.} \bibnamefont{Girvin}},
  \bibnamefont{and} \bibinfo{author}{\bibfnamefont{J.~G.~E.}
  \bibnamefont{Harris}}, \bibinfo{journal}{arXiv:0707.1724}
  (\bibinfo{year}{2007}).

\bibitem[{\citenamefont{Marquardt et~al.}(2007)\citenamefont{Marquardt, Chen,
  Clerk, and Girvin}}]{2007_01_Marquardt_CantileverCooling}
\bibinfo{author}{\bibfnamefont{F.}~\bibnamefont{Marquardt}},
  \bibinfo{author}{\bibfnamefont{J.~P.} \bibnamefont{Chen}},
  \bibinfo{author}{\bibfnamefont{A.~A.} \bibnamefont{Clerk}}, \bibnamefont{and}
  \bibinfo{author}{\bibfnamefont{S.~M.} \bibnamefont{Girvin}},
  \bibinfo{journal}{\prl} \textbf{\bibinfo{volume}{99}},
  \bibinfo{pages}{093902} (\bibinfo{year}{2007}).

\bibitem[{\citenamefont{Wilson-Rae et~al.}(2007)\citenamefont{Wilson-Rae,
  Nooshi, Zwerger, and Kippenberg}}]{2007_02_WilsonRae_Cooling}
\bibinfo{author}{\bibfnamefont{I.}~\bibnamefont{Wilson-Rae}},
  \bibinfo{author}{\bibfnamefont{N.}~\bibnamefont{Nooshi}},
  \bibinfo{author}{\bibfnamefont{W.}~\bibnamefont{Zwerger}}, \bibnamefont{and}
  \bibinfo{author}{\bibfnamefont{T.~J.} \bibnamefont{Kippenberg}},
  \bibinfo{journal}{\prl} \textbf{\bibinfo{volume}{99}},
  \bibinfo{pages}{093901} (\bibinfo{year}{2007}).

\bibitem[{\citenamefont{Braginsky et~al.}(1970)\citenamefont{Braginsky,
  Manukin, and Tikhonov}}]{1970_Braginsky_OpticalCoolingExperiment}
\bibinfo{author}{\bibfnamefont{V.~B.} \bibnamefont{Braginsky}},
  \bibinfo{author}{\bibfnamefont{A.~B.} \bibnamefont{Manukin}},
  \bibnamefont{and} \bibinfo{author}{\bibfnamefont{M.~Y.}
  \bibnamefont{Tikhonov}}, \bibinfo{journal}{Soviet Physics JETP}
  \textbf{\bibinfo{volume}{31}}, \bibinfo{pages}{829} (\bibinfo{year}{1970}).

\bibitem[{\citenamefont{Braginsky et~al.}(2001)\citenamefont{Braginsky,
  Strigin, and Vyatchanin}}]{2001_07_Braginsky_ParametricInstabilityFPCavity}
\bibinfo{author}{\bibfnamefont{V.~B.} \bibnamefont{Braginsky}},
  \bibinfo{author}{\bibfnamefont{S.~E.} \bibnamefont{Strigin}},
  \bibnamefont{and} \bibinfo{author}{\bibfnamefont{S.~P.}
  \bibnamefont{Vyatchanin}}, \bibinfo{journal}{Physics Letters A}
  \textbf{\bibinfo{volume}{287}}, \bibinfo{pages}{331} (\bibinfo{year}{2001}).

\bibitem[{\citenamefont{Kim and Lee}(2002)}]{2002_KimLee_SelfoscillationsAFM}
\bibinfo{author}{\bibfnamefont{K.}~\bibnamefont{Kim}} \bibnamefont{and}
  \bibinfo{author}{\bibfnamefont{S.}~\bibnamefont{Lee}},
  \bibinfo{journal}{Journal of Applied Physics} \textbf{\bibinfo{volume}{91}},
  \bibinfo{pages}{4715} (\bibinfo{year}{2002}).

\bibitem[{\citenamefont{{H\"ohberger} and
  Karrai}(2004)}]{2004_KarraiConstanze_IEEE}
\bibinfo{author}{\bibfnamefont{C.}~\bibnamefont{{H\"ohberger}}}
  \bibnamefont{and} \bibinfo{author}{\bibfnamefont{K.}~\bibnamefont{Karrai}},
  \bibinfo{journal}{Nanotechnology 2004, Proceedings of the 4th IEEE conference
  on nanotechnology} p. \bibinfo{pages}{419} (\bibinfo{year}{2004}).

\bibitem[{\citenamefont{Carmon et~al.}(2005)\citenamefont{Carmon, Rokhsari,
  Yang, Kippenberg, and Vahala}}]{2005_06_Vahala_SelfOscillationsCavity}
\bibinfo{author}{\bibfnamefont{T.}~\bibnamefont{Carmon}},
  \bibinfo{author}{\bibfnamefont{H.}~\bibnamefont{Rokhsari}},
  \bibinfo{author}{\bibfnamefont{L.}~\bibnamefont{Yang}},
  \bibinfo{author}{\bibfnamefont{T.~J.} \bibnamefont{Kippenberg}},
  \bibnamefont{and} \bibinfo{author}{\bibfnamefont{K.~J.}
  \bibnamefont{Vahala}}, \bibinfo{journal}{\prl} \textbf{\bibinfo{volume}{94}},
  \bibinfo{pages}{223902} (\bibinfo{year}{2005}).

\bibitem[{\citenamefont{Kippenberg et~al.}(2005)\citenamefont{Kippenberg,
  Rokhsari, Carmon, Scherer, and Vahala}}]{2005_07_VahalaTheoryPRL}
\bibinfo{author}{\bibfnamefont{T.~J.} \bibnamefont{Kippenberg}},
  \bibinfo{author}{\bibfnamefont{H.}~\bibnamefont{Rokhsari}},
  \bibinfo{author}{\bibfnamefont{T.}~\bibnamefont{Carmon}},
  \bibinfo{author}{\bibfnamefont{A.}~\bibnamefont{Scherer}}, \bibnamefont{and}
  \bibinfo{author}{\bibfnamefont{K.~J.} \bibnamefont{Vahala}},
  \bibinfo{journal}{\prl} \textbf{\bibinfo{volume}{95}},
  \bibinfo{pages}{033901} (\bibinfo{year}{2005}).

\bibitem[{\citenamefont{Marquardt et~al.}(2006)\citenamefont{Marquardt, Harris,
  and Girvin}}]{2005_02_MarquardtHarrisGirvin_Cavity}
\bibinfo{author}{\bibfnamefont{F.}~\bibnamefont{Marquardt}},
  \bibinfo{author}{\bibfnamefont{J.~G.~E.} \bibnamefont{Harris}},
  \bibnamefont{and} \bibinfo{author}{\bibfnamefont{S.~M.}
  \bibnamefont{Girvin}}, \bibinfo{journal}{\prl} \textbf{\bibinfo{volume}{96}},
  \bibinfo{pages}{103901} (\bibinfo{year}{2006}).

\bibitem[{\citenamefont{Corbitt
  et~al.}(2006)}]{2006_08_Corbitt_InstabilityLIGO}
\bibinfo{author}{\bibfnamefont{T.}~\bibnamefont{Corbitt}} \bibnamefont{et~al.},
  \bibinfo{journal}{\pra} \textbf{\bibinfo{volume}{74}},
  \bibinfo{pages}{021802} (\bibinfo{year}{2006}).

\bibitem[{\citenamefont{Carmon and
  Vahala}(2007)}]{2007_CarmonVahala_ModeSpectroscopy}
\bibinfo{author}{\bibfnamefont{T.}~\bibnamefont{Carmon}} \bibnamefont{and}
  \bibinfo{author}{\bibfnamefont{K.~J.} \bibnamefont{Vahala}},
  \bibinfo{journal}{\prl} \textbf{\bibinfo{volume}{98}},
  \bibinfo{pages}{123901} (\bibinfo{year}{2007}).

\bibitem[{\citenamefont{Brown et~al.}(2007)\citenamefont{Brown, Britton,
  Epstein, Chiaverini, Leibfried, and
  Wineland}}]{2007_Wineland_RFcircuitCooling}
\bibinfo{author}{\bibfnamefont{K.~R.} \bibnamefont{Brown}},
  \bibinfo{author}{\bibfnamefont{J.}~\bibnamefont{Britton}},
  \bibinfo{author}{\bibfnamefont{R.~J.} \bibnamefont{Epstein}},
  \bibinfo{author}{\bibfnamefont{J.}~\bibnamefont{Chiaverini}},
  \bibinfo{author}{\bibfnamefont{D.}~\bibnamefont{Leibfried}},
  \bibnamefont{and} \bibinfo{author}{\bibfnamefont{D.~J.}
  \bibnamefont{Wineland}}, \bibinfo{journal}{\prl}
  \textbf{\bibinfo{volume}{99}}, \bibinfo{eid}{137205} (\bibinfo{year}{2007}).

\bibitem[{\citenamefont{Naik et~al.}(2006)}]{2006_08_Schwab_CPB_Molasses}
\bibinfo{author}{\bibfnamefont{A.}~\bibnamefont{Naik}} \bibnamefont{et~al.},
  \bibinfo{journal}{Nature} \textbf{\bibinfo{volume}{443}},
  \bibinfo{pages}{193} (\bibinfo{year}{2006}).

\bibitem[{\citenamefont{Rodrigues et~al.}(2007)\citenamefont{Rodrigues, Imbers,
  Harvey, and Armour}}]{2007_Rodrigues_InstabilitySSET}
\bibinfo{author}{\bibfnamefont{D.~A.} \bibnamefont{Rodrigues}},
  \bibinfo{author}{\bibfnamefont{J.}~\bibnamefont{Imbers}},
  \bibinfo{author}{\bibfnamefont{T.~J.} \bibnamefont{Harvey}},
  \bibnamefont{and} \bibinfo{author}{\bibfnamefont{A.~D.}
  \bibnamefont{Armour}}, \bibinfo{journal}{New Journal of Physics}
  \textbf{\bibinfo{volume}{9}}, \bibinfo{pages}{84} (\bibinfo{year}{2007}).

\bibitem[{\citenamefont{Favero and
  Karrai}(2007)}]{2007_07_FaveroKarrai_ScatteringCooling}
\bibinfo{author}{\bibfnamefont{I.}~\bibnamefont{Favero}} \bibnamefont{and}
  \bibinfo{author}{\bibfnamefont{K.}~\bibnamefont{Karrai}},
  \bibinfo{journal}{arXiv:0707.3117}  (\bibinfo{year}{2007}).

\end{thebibliography}

\end{document}